\documentclass[twocolumn,showpacs,floats,superscriptaddress]{revtex4}

\usepackage[dvips]{graphicx}
\usepackage{amsmath,bm}
\usepackage{psfrag}
\usepackage{dcolumn}
\usepackage{bm}

\def\reff#1{(\ref{#1})}

\usepackage{color}
\definecolor{darkgreen}{rgb}{0,0.5,0}
\definecolor{blue}{rgb}{0,0,0.8}
\definecolor{lightblue}{rgb}{0.93,0.96,1}
\definecolor{darkblue}{rgb}{0.,0.,0.6}
\usepackage[colorlinks,linkcolor=darkgreen,citecolor=darkblue,urlcolor=blue]{hyperref}

\begin{document}

\title{Zigzag transitions and nonequilibrium pattern formation in colloidal chains}

\author{Arthur V. Straube}
\affiliation{Department of Physics, Humboldt University of Berlin, Newtonstr. 15, D-12489 Berlin, Germany}
\affiliation{Rudolf Peierls Centre for Theoretical Physics, University of Oxford, 1 Keble Road, Oxford OX1 3NP, United Kingdom}



\author{Roel P. A. Dullens}
\affiliation{Department of Chemistry, Physical and Theoretical Chemistry Laboratory, University of Oxford, South Parks Road, Oxford, OX1 3QZ, United Kingdom}

\author{Lutz Schimansky-Geier}
\affiliation{Department of Physics, Humboldt University of Berlin, Newtonstr. 15, D-12489 Berlin, Germany}

\author{Ard A. Louis}
\affiliation{Rudolf Peierls Centre for Theoretical Physics, University of Oxford, 1 Keble Road, Oxford OX1 3NP, United Kingdom}

\date{\today}

\begin{abstract}
Paramagnetic colloidal particles that are optically trapped in a linear array can form a zigzag pattern when an external magnetic field induces repulsive interparticle interactions.  When the traps are abruptly turned off, the particles form a nonequilibrium expanding pattern with a zigzag symmetry, even when the strength of the magnetic interaction is weaker than that required to break the linear symmetry of the equilibrium state. We show that the transition to the equilibrium zigzag state is always potentially possible for purely harmonic traps. For anharmonic traps that have a finite height, the equilibrium zigzag state becomes unstable above a critical anharmonicity. A normal mode analysis of the equilibrium line configuration demonstrates that increasing the magnetic field leads to a hardening and softening of the spring constants in the longitudinal and transverse directions, respectively. The mode that first becomes unstable is the mode with the zigzag symmetry, which explains the symmetry of nonequilibrium patterns. Our analytically tractable models help to give further insight into the way that the interplay of such factors as the length of the chain, hydrodynamic interactions, thermal fluctuations affect the formation and evolution of the experimentally observed nonequilibrium patterns.
\end{abstract}

\pacs{82.70.Dd, 47.54.-r}


\maketitle

\section{Introduction}\label{sec:intro}

Confining repelling particles to linear spatially localized traps can lead to the formation of zigzag patterns due to a competition between inter-particle repulsion and the forces of confinement.  Examples include the Frenkel-Kontorova model \cite{Braun-Kivshar-91}, dusty plasmas \cite{Melzer-06}, one-dimensional Wigner \cite{Piacente-etal-prb-04} and ion Coulomb \cite{Pyka-etal-natcomm-13} crystals, colloidal particles \cite{Tatarkova-etal-02, Polin-etal-06, Leonardo-etal-07, Delfau-Coste-Jean-13}, and microfluidic droplet crystals \cite{Beatus-etal-06, Beatus-etal-07}. One particularly fruitful way to study such systems is to use colloidal particles. These offer the possibility to simultaneously visualize and carefully control nonequilibrium behavior using external fields, which facilitates detailed comparisons between experiments and theory~\cite{Loewen-01}. On the one hand, a simple system can be built by confining colloidal particles between two walls or placing them on a spatially confined surface and inducing repulsive interactions between the particles \cite{Chou-Nelson-93, Bongers-Versmold-96, Keim-etal-04}. On the other hand, optical trapping techniques can be used to localize each colloid in an individual optical trap, leading to a one-dimensional colloidal chain of hydrodynamically coupled particles in the form of a line \cite{Tatarkova-etal-02, Polin-etal-06} or a ring \cite{Leonardo-etal-07}. In this study, we focus on the interplay of both these ingredients to create an initially linear chain of colloids.

We have recently shown that repulsively interacting paramagnetic beads, confined by optical traps, can generate nonequilibrium patterns when the traps are abruptly turned off \cite{Straube-etal-11}. The advantage of our system is the general ability to tune interparticle magnetic interactions through the application of an external magnetic field~\cite{Bubeck-etal-99, Tierno-12, Piet-etal-13, Straube-Tierno-13}. At the same time, the colloidal particles are placed in a well-defined initial configuration using optical tweezers. When the traps are abruptly turned off, the long-ranged repulsive intercolloidal interactions generate a dynamically expanding structure that depends on the initial conditions and on the strength of the interactions. The goal of this work is to use detailed calculations and Brownian Dynamics simulations to investigate how an initially one-dimensional configuration of colloids forming a linear chain can generate a dynamically expanding pattern with the observed zigzag symmetry. To understand the equilibrium zigzag transitions and nonequilibrium zigzag patterns, we significantly extend our previously developed theory \cite{Straube-etal-11}. In particular, by obtaining simple analytically tractable models, we gain insight into the way that hydrodynamic interactions and thermal fluctuations affect the nonequilibrium pattern formation. We also show how defects in the zigzag pattern can arise for finite-length chains.

The paper is outlined as follows. We start by describing our system in Sec.~\ref{sec:system}. In Sec.~\ref{sec:eq-zigzag-trans} we focus on the equilibrium zigzag transition and consider the cases of both harmonic and anharmonic traps. In Sec.~\ref{sec:norm-modes} we perform a normal modes analysis which is particularly helpful in clarifying the appearance of zigzag symmetry in nonequilibrium patterns. In Sec.~\ref{sec:dynamic-zigzag} we develop a theory describing nonequilibrium zigzag patterns. Here we first consider an infinite chain of colloids in the presence of no thermal noise and then explore the role and interplay of thermal fluctuations, hydrodynamic interactions, and a finite number of colloids comprising the chain. Our main findings are then summarized in Sec.~\ref{sec:conclusions}.

\section{Chain of optically trapped colloids in magnetic field}\label{sec:system}

Consider a chain of identical paramagnetic colloids trapped optically and subject to a static magnetic field. Experimentally, we use paramagnetic latex spheres of radius $a = 1.35 \; \mu{\rm m}$ (Dynabeads, Invitrogen) immersed in a water solvent filling a $200 \; \mu{\rm m}$ thick quartz glass sample cell. The gravitational length of the particles is much smaller than their sizes and after sedimentation the system becomes effectively two dimensional, with the colloidal configuration in the horizontal plane given by unit vectors $\hat{\mathbf{e}}_1$ and $\hat{\mathbf{e}}_2$. The static spatially uniform magnetic field is applied in the vertical direction, ${\mathbf B}_0=B_0 \hat{\mathbf{e}}_3$. Being paramagnetic, each colloid is polarized along the field and behaves as the induced dipole, whose magnetic moment ${\mathbf m}$ can be approximated by the linear law, ${\mathbf m}=\chi{\mathbf B}_0$, provided that the fields are smaller than the saturation magnetization. Here, $\chi = 3.95 \times 10^{-12} {\rm A \, m^2 T^{-1}}$ is the effective magnetic susceptibility \cite{Blickle-etal-05}.

\begin{figure}[!hbt]
\includegraphics[width=0.47\textwidth]{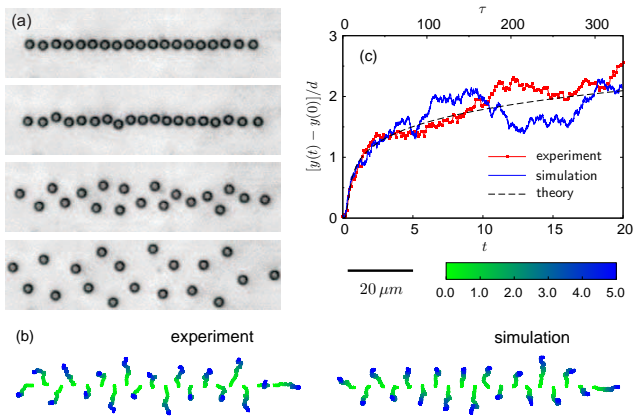}
\caption{(Color online) (a): Microscopy images ($92 \times 23\,\mu$m$^2$) showing an expanding linear chain of $N=19$ magnetic particles with a lattice spacing $d = 4 \,\mu{\rm m}$ upon removing the optical traps at $t=0 \, {\rm s}$ at an external magnetic field $B_0=1.90\,{\rm mT}$ and stiffness $k_0=0.37 \pm 0.01 \,{\rm pN/\mu m}$. The snapshots from left to right are for $t \le 0 \, {\rm s}$, $t=0.2\, {\rm s}$, $t=1 \, {\rm s}$, and $t=5 \, {\rm s}$. (b): The experimental and Brownian dynamics particle trajectories compare well. The simulations are based on Eqs.~\reff{langevin-eq}-\reff{f-s} with the dimensionless parameters $b=0.8 b_c$, $\alpha=30$, and $\sigma=0.001$. The color code indicates the time in seconds. (c): Example of particle trajectories for a single particle in the nonequilibrium pattern, where $y(t)$ is the transverse displacement of the particle. The lines with markers, solid line, and dashed line correspond to the experiment, BD simulations as in panel (b), and the deterministic consideration, see Eq.~\reff{ode-h(t)}, respectively.}
\label{fig:explosion}
\end{figure}

Because of the confinement, ${\mathbf r}\cdot{\mathbf m}=0$ and the field induced by each magnetic dipole is ${\mathbf B}=-\mu_0{\mathbf m}/(4\pi r^3)$, where $\mu_0$ is the permeability of free space and ${\mathbf r}$ is the vector in the plane $(x_1,x_2)$ with the origin at the colloid center. Each magnetic dipole interacts with the external field and the fields induced by all other dipoles. As a result, the energy of interaction $-{\mathbf m(l)}\cdot{\mathbf B}(l')$ of a pair of particles with positions $\mathbf{r}(l)$ and ${\mathbf r}(l')$ corresponds to strictly repulsive interactions described by the potential
\begin{align}
& V_M({\mathbf r}(l),{\mathbf r}(l'))=\frac{\mu_0}{4\pi}\frac{\chi^2 B_0^2}{|{\mathbf r}(l)-{\mathbf r}(l')|^3}\,, \label{repuls-pot}
\end{align}
\noindent where we have omitted the constant contribution caused by the interaction with the external field, $-{\mathbf m}\cdot{\mathbf B}_0$.

Initially, each particle is individually trapped by using optical tweezers. The positions of traps correspond to a one-dimensional array with the spatial period $d$ and are given by ${\mathbf R}(l)=l d\,\hat{\mathbf e}_1$. Although often the optical tweezers are theoretically described as purely harmonic springs, this approximation is valid only close to the trap center. A real trap has a finite range of entrapment, beyond which the particle is practically no longer trapped. For this reason, we model the trapping potential for particle with the position ${\mathbf r}(l)$ via an anharmonic well with a Gaussian profile
\begin{align}
& V_T({\mathbf r}(l),{\mathbf R}(l))=V_0 -V_0\exp{\left[-\frac{1}{2}\frac{k_0}{V_0}\delta \mathbf{r}^2(l)\right]}\, , \label{gauss-trap-pot}
\end{align}
\noindent which is consistent with recent measurements of the optical trapping potential \cite{Juniper-etal-12}. Here, $\mathbf{\delta r}(l) ={\mathbf r}(l)-{\mathbf R}(l)$ and the additive constant is chosen such that the minimum corresponds to $V_T=0$. The parameters $V_0$ and $k_0$ specify, respectively, the the depth of the well and the stiffness of the potential. The anharmonic nature or softness of the trap is characterized by the dimensionless parameter
\begin{equation}
\alpha = \frac{k_0 d^2}{V_0}\,. \label{alpha-def}
\end{equation}
\noindent Rewriting Eq.~\reff{gauss-trap-pot} in dimensionless form, we obtain $V_T/(k_0 d^2)=\alpha^{-1}-\alpha^{-1}\exp[-\alpha\, \delta \mathbf{r}^2/(2d^2)]$ with the effective range of attraction $\sqrt{2/\alpha}$. We see that the limit of $\alpha \to 0$ of expression \reff{gauss-trap-pot} corresponds to the purely harmonic potential, $V_T({\mathbf r}(l),{\mathbf R}(l))=k_0 \delta r^2(l)/2$, for all $\delta r(l)$. For $\alpha > 0$ the trapping is nearly quadratic for small $\delta r(l)$ but becomes increasingly anharmonic at larger $\delta r(l)$ and has a finite height $V_0$. Above this barrier the particle can escape from the trap. It is important to note that $k_0$ and $V_0$ can be changed in the experiment but that their ratio always remains virtually constant and therefore characterizes the optical trap \cite{Juniper-etal-12}.

Based on these basic ingredients, the repulsive, Eq.~(\ref{repuls-pot}), and optical trap, Eq.~(\ref{gauss-trap-pot}), potentials, the total energy of the crystal can be written as a superposition
\begin{align}
& U=\frac{1}{2}\sum_l\sum_{l'\ne l} V_{M}({\mathbf r}(l),{\mathbf r}(l')) +\sum_l V_{T}({\mathbf r}(l),{\mathbf R}(l))\,, \label{tot-energy}
\end{align}
\noindent in which we ignore hard-core repulsive interactions. Note that the trap potential plays the role of a restoring force that tends to hold a given particle at a prescribed position, whereas the force exerted on the particle due to magnetic interactions tends to push the particle away from this position. Depending on the relative strengths of these two competing tendencies, different equilibriums states are possible. One expects that as the magnetic field is increased the initial line configuration is broken and the transition to an equilibrium zigzag state takes place, which is discussed next.

\section{Equilibrium zigzag transition}\label{sec:eq-zigzag-trans}

\subsection{Interaction potential for zigzag configuration}\label{ssec:eq-zigzag-potential}

We now analyze properties of the equilibrium zigzag state for the simplest situation with the total number of particles $N \to \infty$ and no thermal noise; the impact of finite temperature is discussed in Sec.~\ref{ssec:eq-zigzag-th-noise}. As the magnetic energy remains invariant under translation, the equilibrium zigzag configuration can be presented via
\begin{align}
\frac{\mathbf{h}(l)}{d} = l \, \hat{\mathbf e}_1+ (-1)^l \frac{h}{2} \,\hat{\mathbf e}_2 \label{zigstate-h}
\end{align}
\noindent with $l=0, \pm 1, \pm 2, \dots, \pm M$, $M=(N-1)/2 \to \infty$. Here, $\mathbf{h}(l)$ is the equilibrium displacement of particle $l$ from the center of its trap and $h$ is the dimensionless order parameter characterizing the transition to the zigzag state. The state with $h=0$ corresponds to the trivial line state, while the ground state with $h \ne 0$ describes the zigzag configuration.

The total equilibrium energy $U^{eq}$ of the colloidal configuration is given by Eq.~\reff{tot-energy} with ${\mathbf r}(l)={\mathbf h}(l)$. Because the chain of colloids under consideration is infinite, each particle makes an identical contribution $U_0$ into the total energy $U^{eq}$. As a result, we simplify our analysis by studying the averaged energy per particle, $U_0 =\lim_{N \to \infty} U^{eq}/N$. It follows from Eq.~\reff{tot-energy} that $U_0(h)=(1/2)\sum_{l' \ne l} V_M ({\mathbf h}(l),{\mathbf h}(l'))+V_T({\mathbf h}(l),{\mathbf R}(l))$, where parts of the sum can be evaluated analytically, see Refs.~\cite{Kantorovich-etal-09, Kantorovich-etal-12}. Representing the sum over $l'$ as two sums over $m=1, 2, \dots M$ ($M \to \infty$) with $l'=l-m$ and $l'=l+m$ and noticing that the summands with the same $m$ are equal, we arrive at
\begin{align}
\frac{U_0(h)}{k_0L^2}&=b^2\sum_{m=1}^{\infty} f_m^{3}(h)+\frac{1-g(h)}{\alpha}\,, \label{zigzag-U0}\\
f_m(h)&=\frac{1}{\sqrt{m^2+p_m h^2}}, \quad p_m=\frac{1-(-1)^m}{2}\,, \label{fm-pm-def}\\
g(h)&=\exp\left(-\frac{1}{8}\alpha h^2\right)\,, \\
b^2 & = \frac{\mu_0}{4\pi}\frac{\chi^2 B_0^2}{k_0 d^5}\,. \label{b2-def}
\end{align}
\noindent The first term in potential (\ref{zigzag-U0}) describes repulsive interactions between a given particle with all its neighbors, as accounted by the summation index $m$: $m=1$ corresponds to the nearest neighbors, $m=2$ to the next nearest neighbors, and so on. The second term describes the optical trapping with the dimensionless parameter $\alpha$ characterizing the softness of the trap, as given by  Eq.~(\ref{alpha-def}). The dimensionless parameter $b$ introduced by Eq.~(\ref{b2-def}) stands for the intensity of the magnetic field relative to the characteristic energy of trapping, $k_0 d^2$.

For convenience, we renormalize potential (\ref{zigzag-U0}) such that $U_0(0)=0$ ($h=0$ means the line state) and hereafter measure it in the units of $k_0 d^2$. As a result, for the rescaled potential we have
\begin{equation}
U_0(h)=b^2\left[\sum_{m=1}^{\infty} f_m^{3}(h) -\zeta(3) \right]+\frac{1-g(h)}{\alpha}, \label{poten-U0}
\end{equation}
\noindent where we have introduced the Riemann Zeta function $\zeta(x)=\sum_{m=1}^{\infty}m^{-x}$.

The infinite sum in relation \reff{poten-U0} can be evaluated only numerically, which significantly restricts the possibility of analytic analysis. Quite a helpful simplification of Eq.~\reff{poten-U0} is the nearest-neighbor (NN) approximation. Neglecting all the terms except for that one with $m=1$ in Eq.~\reff{zigzag-U0} and again renormalizing the potential such that it vanishes at $h=0$, in the NN approximation we obtain
\begin{equation}
U_0^{NN}(h)=b^2\left[ \frac{1}{(1+h^2)^{3/2}} - 1 \right]+\frac{1-g(h)}{\alpha}. \label{poten-U0-NNA}
\end{equation}
\noindent As we see below, this approximation works very well. Next, we consider the case of a purely harmonic trap, $\alpha=0$, and then discuss how the softness of the trap affects the results.

\subsection{Harmonic trap}\label{ssec:eq-zigzag-harmonic}

In the limiting case of harmonic trap, $\alpha=0$, the trapping term $\alpha^{-1}[1-g(h)]$ in Eqs.~(\ref{poten-U0}) and (\ref{poten-U0-NNA}) reduces to a purely quadratic contribution, $(1/8)h^2$. The requirement of energy minimum, which is given by the conditions $U^{\prime}_0(h_{*})=0$ and $U^{\prime\prime}_0(h_{*})>0$, determines the stable equilibrium solution as a function of the field, $h_{*}=h_{*}(b)$. Hereafter primes abbreviate the derivatives with respect to $h$. The condition $U^{\prime}_0(h_{*})=0$, admits the trivial line state, $h_{*}=0$, and a nontrivial zigzag state with $h_{*} \ne 0$, obeying an equation
\begin{equation}
-b^2\sum_{m=1}^{\infty} p_m f_m^{5}(h_{*}) +\frac{1}{12}=0\,. \label{min-U0-cond-harm}
\end{equation}
\noindent We expect that the line state, $h_{*}=0$, is stable at fields $b<b_c$ and the zigzag transition occurs at some critical field $b=b_c$.
Accordingly, the threshold value $b_c$ is obtained from Eq.~(\ref{min-U0-cond-harm}) at $h_{*}=0$, which gives
\begin{equation}
b_{c}=\sqrt{\frac{8}{93\, \zeta(5)}}\approx 0.288\, \label{bc-gen}
\end{equation}
\noindent as the maximum field at which $h_{*}$ is a stable solution. Here, we have taken into account that $\sum_{m=1}^{\infty}p_m m^{-5}=\sum_{n=1}^{\infty}(2n-1)^{-5}=(31/32)\zeta(5)$.

Since Eq.~(\ref{min-U0-cond-harm}) admits no analytical solution for $h_{*}(b)\ne 0$, we consider two further approximations below.

\subsubsection{Analytical solution close to critical point}\label{ssec:}

Here we analyze potential (\ref{poten-U0}) for the case where $\alpha=0$ and the system is close to the critical point, where $h$ is small. By expanding the sum as a series with respect to $h$
\begin{equation}
\sum_{m=1}^{\infty} f_m^{3}=\zeta(3)-\frac{93}{64}\zeta(5) h^2+ \frac{1905}{1024}\zeta(7) h^4+\mathcal{O}(h^6)\, \label{} \nonumber
\end{equation}
\noindent we obtain for the potential
\begin{align}
U_0(h)=\beta(b) h^2+\gamma(b) h^4+\mathcal{O}(h^6)\, \label{U0-expan-harm}
\end{align}
with
\begin{align}
\beta(b)=\frac{1}{8}-\frac{93}{64}\zeta(5)b^2, \quad \gamma(b)=\frac{1905}{1024}\zeta(7)b^2.  \nonumber 
\end{align}
\noindent In the case of no field, $b=0$, the potential $U_0 \propto h^2$ with a coefficient $\beta(0)>0$, as expected. So, $U_0$ has
a global minimum at $h=h_{*}=0$, which corresponds to the trivial ground state in the form of line. As the field $b$ is increased, $\beta(b)$ decreases and becomes negative. As $\gamma >0$, the fact that $\beta(b)<0$ implies that the global minimum is at a nontrivial $h=h_{*}\ne 0$, which corresponds to the zigzag state. The onset of the zigzag state occurs when $\beta(b_c)=0$, providing the same value of $b_c$ as in Eq.~(\ref{bc-gen}).

Rewriting potential (\ref{U0-expan-harm}) close to the threshold, where $h \ll 1$ and $b = b_c + \delta b$, $|\delta b| \ll b_c$, we end up with
\begin{equation}
U_0(h) \approx \frac{1}{8}\left(\frac{b_c^2-b^2}{b_c^2}\right) h^2+\gamma(b_c) h^4.\label{U0-small-h-harm}
\end{equation}
\noindent As follows from Eq.~(\ref{U0-small-h-harm}), for the line state, $h_{*}=0$, we have $U_0^{\prime\prime}(0)=(b_c^2-b^2)/(4b_c^2)$, which has to be positive for stability. Thus, the line state is stable for $b<b_c$, as expected. To obtain the solution beyond the critical point, we minimize potential (\ref{U0-small-h-harm}) for $h_{*}\ne 0$. As a result, we arrive at a Landau-like square-root law describing the zigzag state
\begin{equation}
h_{*}=\pm \sqrt{C(b^2-b_c^2)}, \;\;  b>b_c  \label{h(b)-small-h-harm}
\end{equation}
\noindent with the constant $C=(64)/[1905\,b_c^4 \, \zeta(7)]$, see also Fig.~\ref{fig:bif_diag-harm}. We note that the same expression can be obtained by simplifying  Eq.~\reff{min-U0-cond-harm} directly close to the threshold. For the zigzag state, we find that $U_0^{\prime\prime}(h_{*})=(b^2-b_c^2)/(2b_c^2)$, indicating that it is stable for $b > b_c$. We also note that a similar square-root buckling singularity was pointed out for a chain of particles with screened electrostatic interactions \cite{Delfau-Coste-Jean-13} and for a system of multilayered crystalline sheets of macroions under slit confinement \cite{Oguez-etal-09}.

\subsubsection{Nearest neighbor approximation}\label{ssec:}

Another possibility that admits analytical analysis is the NN approximation. As earlier, the line state is stable below the critical point. The stable solution at supercritical conditions can be obtained from Eq.~(\ref{min-U0-cond-harm}), where only the leading term, $m=1$, is retained in the sum
\begin{equation}
-b^2\left(1+h_{*}^2\right)^{-5/2}+\frac{1}{12}=0\,. \label{min-U0-cond-harm-nna}
\end{equation}
\noindent By putting $h_{*}=0$ in Eq.~(\ref{min-U0-cond-harm-nna}), we find the critical field
\begin{equation}
b^{NN}_c=\frac{1}{\sqrt{12}}\approx 0.289\,, \label{bc-nna}
\end{equation}
\noindent which is very close to the general result, cf. Eq.~(\ref{bc-gen}). Taking into account Eq.~(\ref{bc-nna}), it follows from Eq.~(\ref{min-U0-cond-harm-nna}) that
\begin{equation}
h_{*}^{NN}= \pm b_c^{-2/5}\sqrt{b^{4/5}-b_c^{4/5}}\,, \;\; b>b_c,  \label{h(b)-harm-nna}
\end{equation}
\noindent where $b_c$ is defined by expression \reff{bc-nna}. Note that in contrast to result (\ref{h(b)-small-h-harm}), solution (\ref{h(b)-harm-nna}) is valid not only close to the threshold but at any value of $b>b_c$.

\subsubsection{Comparison of results}\label{ssec:}

\begin{figure}[!h]
\includegraphics[width=0.45\textwidth]{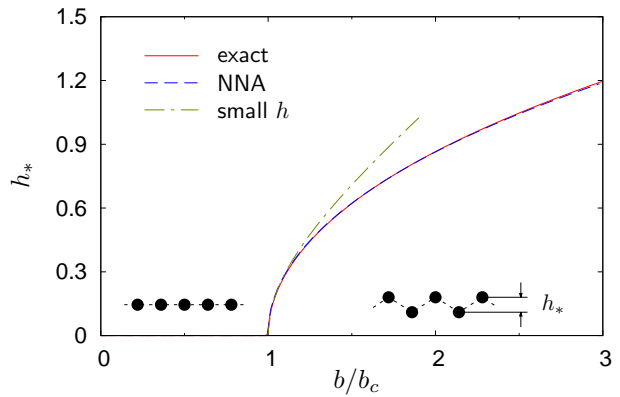}
\caption{(Color online) Diagram showing the transition from the line ($h_{*}=0$) to the zigzag ($h_{*}\ne 0$) state for the harmonic trap, $\alpha=0$, as the field $b$ is increased. The exact solution is according to Eq.~(\ref{min-U0-cond-harm}). The approximate solutions for small $h$ (dash dotted line) and in the NN-approximation (dashed line) are described by formulas (\ref{h(b)-small-h-harm}) and (\ref{h(b)-harm-nna}), respectively. Note that the exact solution and that in the NN-approximation are practically indistinguishable in the scale of the figure.} \label{fig:bif_diag-harm}
\end{figure}

Finally, we solved Eq.~(\ref{min-U0-cond-harm}) numerically and determined the corresponding dependence $h_{*}(b)$. The results are illustrated in Fig.~\ref{fig:bif_diag-harm}, where we also perform the comparison with the approximate solutions, Eqs.~(\ref{h(b)-small-h-harm}) and (\ref{h(b)-harm-nna}).

As expected, the solution valid for weak supercriticality, Eqs.~(\ref{h(b)-small-h-harm}), works well in the vicinity of the critical point $b=b_c$ and starts to deviate for larger fields. We also emphasize the impressively good agreement of the exact solution with that in the NN approximation, Eq.~(\ref{h(b)-harm-nna}). Physically, this finding indicates that the long-range nature of the repulsive interactions is not important: each colloid in the ground state interacts with its neighbors only, while the contribution made by interactions with the next neighbors is nearly vanishing. Mathematically, this point is ensured by a rather fast convergence of the sum in Eqs.~(\ref{poten-U0}) and (\ref{min-U0-cond-harm}).

\subsection{Anharmonic trap}\label{ssec:eq-zigzag-nonharmonic}

We now proceed to the discussion of how the anharmonicity of the trap affects the equilibrium states. The trivial line state, $h_{*}=0$, is independent of the softness parameter and for $\alpha>0$ remains stable for subcritical fields, $b<b_c$. To investigate the effect of anharmonic trap on the nontrivial state, $h_{*}\ne 0$, Eq.~(\ref{min-U0-cond-harm}) should be replaced by
\begin{equation}
-b^2\sum_{m=1}^{\infty} p_m f_m^{5}(h_{*}) +\frac{1}{12}\exp\left(-\frac{1}{8}\alpha h_{*}^2\right)=0\,, \label{min-U0-cond}
\end{equation}
\noindent which determines extrema of potential $U_0(h)$ for arbitrary $\alpha$ and allows to obtain the dependence $h_{*}=h_{*}(b,\alpha)$. The critical value $b_c$ follows from Eq.~(\ref{min-U0-cond}) at $h_{*}=0$, which does not differ from the similar condition in the case of $\alpha=0$, see Sec.~\ref{ssec:eq-zigzag-harmonic}. As a result, independent of $\alpha$, the value $b_c$ is given by expression (\ref{bc-gen}), or by formula (\ref{bc-nna}) in the NN approximation, and the line state, $h_*=0$, is stable at any $b < b_c$, and for any $\alpha$.

\subsubsection{Analytical solution close to critical point}\label{ssec:}

We note that in the case of arbitrary $\alpha$, Eq.~(\ref{min-U0-cond}) admits no analytic solution, even in the NN approximation. However, the dependence $h_{*}(b)$ can be obtained close to the threshold, where $h \ll 1$ and $|b-b_c| \ll b_c$. The representation of potential (\ref{U0-small-h-harm}) remains formally similar, though with a modified coefficient
\begin{equation}
U_0(h) \approx \frac{1}{8}\left(\frac{b_c^2-b^2}{b_c^2}\right) h^2+\Gamma \, h^4, \;\; \Gamma=\gamma(b_c)-\frac{\alpha}{128}\,. \label{U0-small-h}
\end{equation}
\noindent Note that in contrast to $\gamma(b_c)>0$ in Eq.~(\ref{U0-small-h-harm}), the coefficient $\Gamma=\Gamma(\alpha)$ can change the sign depending on $\alpha$. The border value of $\alpha_c$ is determined by the condition $\Gamma(\alpha_c)=0$, which yields
\begin{equation}
\alpha_c = 128 \gamma(b_c)=\frac{635 \,\zeta(7)}{31 \,\zeta(5)}\approx 19.9\,. \label{alphac-gen}
\end{equation}
\noindent Thus, for traps with $\alpha < \alpha_c$, $\Gamma>0$ and the situation is qualitatively similar to the case of harmonic trap. For traps with $\alpha > \alpha_c$, $\Gamma<0$ and the solution becomes drastically different.

To interpret this difference, consider the nontrivial solution of equation $U_0^{\prime}(h_{*})=0$ with $U_0$ given by Eq.~(\ref{U0-small-h}). Close to the threshold we obtain
\begin{equation}
h_{*} = \pm \sqrt{\frac{8(b^2-b_c^2)}{b_c^2(\alpha_c-\alpha)}}\,, \label{h(b)-small-h}
\end{equation}
\noindent which coincides with Eq.~(\ref{h(b)-small-h-harm}) in the case of $\alpha=0$.

As follows from Eq.~(\ref{h(b)-small-h}), $h_{*}(b)$ demonstrates a supercritical pitchfork bifurcation as a function of $b$ for $\alpha<\alpha_c$, which is stable, as shown earlier, in Sec.~\ref{ssec:eq-zigzag-harmonic}. It shows a subcritical pitchfork bifurcation for $\alpha > \alpha_c$, see also Fig.~\ref{fig:bif_diag}. Because $U_0^{\prime\prime}(h_{*})=(b^2-b_c^2)/(2b_c^2)$, the latter solution corresponds to the maximum of potential and is therefore unstable. As a result, we can conclude that for traps characterized by $\alpha>\alpha_c$ the transition to the zigzag state is impossible in principle. For such traps, the line state is stable for $b<b_c$. Beyond the threshold value the particles escape from the traps before the zigzag state can form, which corresponds to a dynamical expansion of configuration \cite{Straube-etal-11}, with formally $h \to \infty$.

We also note that Eqs.~(\ref{U0-small-h}) and (\ref{h(b)-small-h}) are formally valid in the NN approximation, in which $b_c$ is given by Eq.~(\ref{bc-nna}), $\gamma^{NN}(b_c)=(15/8)b_c^2$, and $\alpha_c^{NN}=20$. The latter is again very close to result (\ref{alphac-gen}).

\subsubsection{Equilibriums states. General picture}\label{ssec:eq-zigzag-gen}

\begin{figure}[!h]
\includegraphics[width=0.45\textwidth]{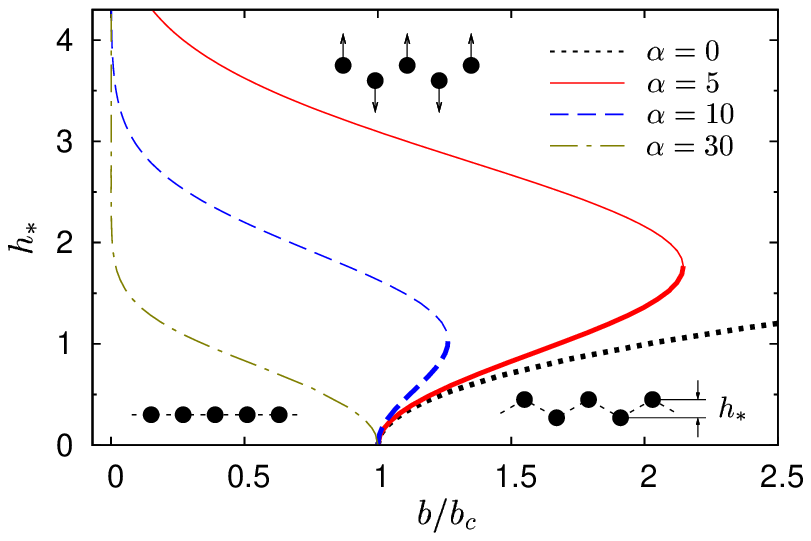}
\caption{(Color online) Bifurcation diagram showing equilibrium states for different values of $\alpha$. The line state $h_{*}=0$ exists for all $\alpha$ and is stable for $b<b_c$. The solution describing the stable equilibrium zigzag state ($db/dh_{*}>0$) is shown by bold lines; it exists for traps with $\alpha < \alpha_c$ in the range of $b_c<b<b_{**}$ with $b_{**}$ the turning point, see Eq.~(\ref{b**-nna}). The branches where $db/dh_{*}<0$ (lines of normal thickness) are unstable.} \label{fig:bif_diag}
\end{figure}

To obtain the complete picture of equilibriums states for arbitrary $h_{*}$, we solved Eq.~(\ref{min-U0-cond}) numerically. This equation corresponds to the requirement $U_0^{\prime}(h_{*})=0$ and determines extremal solutions $h_{*} \ne 0$. The corresponding solutions $h_{*}$ as functions of $b$ for different $\alpha$ are presented on the in Fig.~\ref{fig:bif_diag}. To elucidate which of the branches are stable we provide a general requirement necessary for stability
\begin{equation}
U_0^{\prime\prime}(h_{*})=\frac{h_{*}}{2b}\left(\frac{db}{dh_{*}}\right)\exp\left(-\frac{\alpha}{8}h_{*}^2\right) > 0\,, \label{d2U0-stab-gen}
\end{equation}
\noindent which is obtained by differentiating potential (\ref{poten-U0}) and using Eq.~(\ref{min-U0-cond}). As $\alpha$, $b$ and $h_{*}$ are positive, the stability is determined by the sign of the derivative $db/dh_{*}$.

We now interpret the dependencies $h_{*}(b)$ starting at vanishing $b$, where the line state ($h_{*}=0$) is stable, and then gradually increasing the field, $b$. The numerical results confirm the main conclusion following from formula (\ref{h(b)-small-h}). The critical value $\alpha$ separates two classes of systems that exhibit qualitatively different behavior.

Consider first the case of relatively stiff traps characterized by $\alpha < \alpha_c$. The trivial line state ($h_{*}=0$, $b<b_c$) bifurcates at $b=b_c$ into a zigzag state via the pitchfork bifurcation so that $h_{*} \ne 0$ at $ b>b_c$. As follows from Eq.~(\ref{d2U0-stab-gen}), the zigzag state is stable since $U_0^{\prime\prime}(h_{*})\propto (db/dh_{*})>0$. Note that in the limiting case of harmonic trap, $\alpha=0$, the transition to the stable zigzag state exists for any $b>b_c$, which is in agreement with the results in Sec.~\ref{ssec:eq-zigzag-harmonic}. For $0<\alpha<\alpha_c$, we observe a new feature caused by softness of the trapping potential, which is not captured by asymptotic solution (\ref{h(b)-small-h}). By increasing $b$ further we reach a turning point ($b_{**}, h_{**}$), at which $U_0^{\prime\prime}(h_{*}) = (db/dh_{*})=0$. This behavior of $h_{*}(b)$ corresponds to a saddle-node bifurcation. In the NN approximation, we obtain analytic expressions for the turning point
\begin{eqnarray}
b_{**}^{NN}&=&b_c^{NN}\left(\frac{\alpha_c^{NN}}{\alpha}\right)^{5/4}\exp\left(-\frac{\alpha_c^{NN}-\alpha}{16}\right)\,, \label{b**-nna} \\
h_{**}^{NN}&=&\pm \sqrt{\frac{\alpha_c^{NN}-\alpha}{\alpha}} \quad (\alpha < \alpha_c)\,. \label{}
\end{eqnarray}

Beyond the field $b_{**}$, no equilibrium solutions with finite $h_{*}$ exist. The remaining part of the branch with finite $h_{*}$ in the range $b<b_{**}$, $h_{*}>h_{**}$, is characterized by $U_0^{\prime\prime}(h_{*})\propto (db/dh_{*})<0$, and is therefore unstable. This solution corresponds to the maximum of the potential and sets the potential barrier separating the domains of the equilibrium zigzag state and the formal solution with $h_{*} \to \infty$. The latter solution means that all the particles escape from their traps and move apart due to repulsion. Thus, for the traps with $\alpha < \alpha_c$ the stable zigzag state exists in the finite range of fields, $b_c<b<b_{**}$.

In summary, for softer traps, $\alpha > \alpha_c$, there is no stable zigzag state. Instead, for $b < b_c$ the fixed point is the line, and for $b>b_c$ the fixed point is a solution with $h_{*} \to \infty$. For $\alpha < \alpha_c$, the line is still stable for $b<b_c$. For $b_c < b < b_{**}$, a stable zigzag state is possible, and for $b > b_{**}$ the only solution is that with $h_{*} \to \infty$.

Thus, the softness of the trap potential, $\alpha$, determines whether an equilibrium zigzag transition can be observed. We recall that up to now we have considered the case of zero temperature. In a real system, one has to take into account thermal fluctuations. This issue is discussed in Sec.~\ref{ssec:eq-zigzag-th-noise}.

\subsection{Impact of thermal noise}\label{ssec:eq-zigzag-th-noise}

Here we briefly discuss how the critical values $b_c$ and $b_{**}$ are affected by the temperature. First of all, it is important to note that in contrast to purely one-dimensional systems, in which phase transitions may not exist due to thermal noise \cite{Kastner-RMP-08}, our system is a two-dimensional (or only a quasi-one dimensional) system and thus can exhibit a true phase transition. Then, consider the case of a purely harmonic trap, $\alpha=0$, in which at zero temperature we have a square-root singularity, as given by Eq.~\reff{h(b)-small-h-harm}. Generally, the sharp deterministic threshold is known to be blurred by thermal fluctuations, leading to a smooth transition regime for the noisy system \cite{Meunier-Verga-88}; see also Ref.~\cite{Delfau-Coste-Jean-13}, presenting a recent detailed study of the noisy zigzag transition. At any nonzero temperature, the singularity at the threshold is replaced by a bifurcation region, whose range depends on the thermal energy $k_BT$. Outside this region, the system is strongly stable and the thermal fluctuations do not modify the stability of the system, which is in either the linear (with the averaged transverse displacement $\left<\delta y \right>=0$) or a single zigzag (with $\left<\delta y \right>=h_*$ or $\left<\delta y \right>=-h_*$) configuration. Within the bifurcation region, the system is very sensitive to any small perturbation and it randomly flips between the two symmetric zigzag states with $\left<\delta y \right>=h_*$ and $\left<\delta y \right> =-h_*$. Unlike in Ref.~\cite{Delfau-Coste-Jean-13}, our system is additionally characterized by an anharmonic trapping potential, $\alpha>0$, meaning that thermal noise can lead to the instability of the equilibrium configurations and trigger dynamically expanding zigzag patterns at fields effectively lower than those prescribed by the deterministic values $b_c$ and $b_{**}$.

Consider first the case of $\alpha > \alpha_c$, when there is no stable zigzag state. The potential barrier $\Delta U = U_0(h_{*})-U_0(0)$ can be evaluated close to the threshold $b=b_c$. Note that in the case under consideration, formula (\ref{h(b)-small-h}) provides the solution corresponding to the maximum of potential. Taking into account that $U_0(0)=0$ and substituting Eq.~(\ref{h(b)-small-h}) into Eq.~(\ref{U0-small-h}), we evaluate the height of potential. Since we measure energy in the scales of $k_0 d$, we have
\begin{equation}
\frac{\Delta U}{k_BT} = \frac{k_0d^2}{k_BT}\, \frac{\left(b_c^2-b^2\right)^2}{2b_c^2(\alpha-\alpha_c)}\,. \nonumber\label{deltaU-bc}
\end{equation}
\noindent Solving for $b_c^{\prime}$ which holds when $\Delta U/(k_BT)=1$, provides an estimate for the critical field at finite temperature
\begin{equation}
b_c^{\prime} \approx b_c \left[1- \sqrt{\frac{(\alpha-\alpha_c)}{2\alpha}\frac{k_BT}{V_0}}\right]\,,\label{bc-T-shift}
\end{equation}
\noindent where $b_c$ has the meaning of the critical field at $k_BT=0$. From Eq.~\reff{bc-T-shift} we see that at vanishing temperature, the critical field $b_c^{\prime}=b_c$, as expected. At nonzero temperature, however, the critical field becomes smaller than $b_c$, because thermal fluctuations allow the particles to escape from the potential well. Note that this shift is determined by the ratio of thermal energy to the characteristic depth of the well $V_0$.

We now obtain a similar estimate for the case of explosion from the zigzag state, $h_{*}\ne 0$, which is only possible if $\alpha<\alpha_c$. At zero temperature this happens when $b=b_{**}$ and $h_{*}=h_{**}$. To evaluate the height of potential barrier we first find $h_{*}$ in the vicinity of the turning point ($b_{**}, h_{**}$), which can be done in the NN approximation
\begin{equation}
h_{*}=h_{**}+\delta h, \quad \delta h = \pm \frac{\alpha_c}{\alpha} \sqrt{\frac{\alpha(b_{**}^2-b^2)}{5 \,b_{**}^2(\alpha_c-\alpha)}}\,. \nonumber \label{}
\end{equation}
\noindent Using this result, we evaluate the potential barrier $\Delta U=U_0(h_{**}+\delta h)-U_0(h_{**}-\delta h)$
\begin{equation}
\frac{\Delta U}{k_BT} = \frac{k_0d^2}{k_BT} \, \frac{5\alpha_c}{3\alpha}\exp\left(\frac{\alpha-\alpha_c}{8}\right) \left(\frac{b_{**}^2-b^2}{5b_{**}^2}\right)^{3/2}  \nonumber \label{deltaU-bc}
\end{equation}
\noindent and finally arrive at an estimate for the critical field at finite temperature
\begin{equation}
b_{**}^{\prime} \approx b_c \left[1- \frac{1}{40}\exp\left(\frac{\alpha_c-\alpha}{12}\right)\left(30 \, \frac{k_BT}{V_0}\right)^{2/3}\right].\label{}
\end{equation}
Similar to expression (\ref{bc-T-shift}), we have $b_{**}^{\prime}=b_{**}$ at $T=0$ and a reduction in the field due to thermal fluctuations that depends on the dimensionless parameter $k_BT/V_0$.

\section{Normal mode analysis}\label{sec:norm-modes}

\subsection{Infinite chain, no thermal noise}\label{ssec:norm-modes-analytics}

We now use a normal mode analysis to investigate the phonon dispersion relations. We start with the simplest situation of an infinite chain and no thermal noise. We represent the instant particle positions as ${\mathbf r}(l)={\mathbf R}(l)+{\mathbf u}(l)$, where ${\mathbf u}(l)$ is a perturbation describing small deviations from the equilibrium line state, $h=0$. The total energy of the colloidal crystal \reff{tot-energy} is expanded to lowest order around this equilibrium state to give $U = U^{eq} + U^{harm}$, where $U^{eq}(0)$ is independent of perturbations and is therefore irrelevant for the normal mode analysis. For the correction caused by the perturbation, in the harmonic approximation we generally have \cite{Ashcroft-Mermin, Gruenberg-Baumgartl-07}, see also Ref.~\cite{Chou-Nelson-93}:
\begin{align}
& U^{harm}=\frac{1}{2}\,\sum_{l l'}\sum_{\mu, \nu} u_{\mu}(l) \Phi_{\mu \nu}(l l') u_{\nu}(l')\,, \label{U-harm-gen}\\
& \Phi_{\mu \nu}(l l') = \left.\frac{\partial^2 U}{\partial x_\mu(l) \partial x_\nu(l')} \right|_{\rm eq}\,, \label{} \nonumber
\end{align}
\noindent where $\mu, \nu \in \{1,2\}$ and the elements $\Phi_{\mu \nu}(l l')$ of the Hessian matrix are taken at the equilibrium, ${\mathbf r}(l)={\mathbf R}(l)$ for all $l$. Because the total energy  $U$ comprises the contributions caused by the optical trapping and magnetic repulsive interactions, we similarly have $\Phi_{\mu \nu}(l l') = \Phi_{\mu \nu}^{T}(l l')+\Phi_{\mu \nu}^{M}(l l')$, where
\begin{align}
& \Phi_{\mu \nu}^{T}(l l') = \delta_{\mu\nu} \delta_{ll'}\,,  \nonumber\\
& \Phi_{\mu \nu}^{M}(l l') = b^2 \delta_{\mu\nu} \left( \delta_{ll'} \sum_{l''\ne l}\frac{C_{\mu}}{|l-l''|^5}-\frac{C_{\mu}}{|l-l'|^5}\right)\,, \label{} \nonumber
\end{align}
\noindent with the coefficients
\begin{align}
&  C^{||} \equiv C_1 = 12, \quad C^{\perp} \equiv C_2 = -3\,.  \label{} \nonumber
\end{align}

In contrast to studies in Refs.~\cite{Pyka-etal-natcomm-13, Delfau-Coste-Jean-13}, the dynamics of perturbations in our system is overdamped \cite{Straube-etal-11, Ohshima-Nishio-01} due to the viscous fluid and is described by equations
\begin{align}
&\dot{u}_{\mu}(l)=-\frac{\partial U^{harm}}{\partial u_{\mu}(l)}=-\sum_{l'} \sum_{\nu} \Phi_{\mu \nu}(l l')u_{\nu}(l') \,, \label{NM-eq-pert}
\end{align}
\noindent where the friction coefficient $\xi=6\pi\eta a$ between the particles and the solvent with the dynamic viscosity $\eta$ is absorbed in the time units. To introduce dimensionless variables, the length and energy are measured respectively in the units of $d$ and $k_0 d^2$, as before; the time is expressed in the scale of $\xi/k_0$.

Rearranging the summation in Eq.~\reff{NM-eq-pert} such that $l'=l+m$ with $m$ looping over the neighbors of particle $l$ and using the symmetry $\Phi_{\mu\nu}(ll')=\Phi_{\mu\nu}(l'l)$, we arrive at a set of coupled differential equations
\begin{align}
&\dot{u}_{\mu}(l) =  \mathcal L_{\mu} u_{\mu}(l) \,, \label{NM-eq-umu} \\
& \mathcal L_{\mu}= -1 + b^2 C_{\mu} \sum_{m=1}^{\infty} \frac{E^{-m}-2+E^{m}}{m^5}\,. \label{LinOper-L}
\end{align}
\noindent The first term in the linear operator $\mathcal L_{\mu}$ comes from the optical trapping, while the contribution $\propto b^2$ is caused by repulsive interactions. In the latter, $m$ accounts for the magnetic interaction of particle $l$ with its the nearest neighbors ($m=1$), next nearest neighbors ($m=2$), and so on. For compactness, we have introduced a shift operator $E^{\pm m}$ that acts such that $E^{\pm m} u_{\mu}(l)=u_{\mu}(l \pm m)$.

The normal modes are readily found by means of the ansatz
\begin{align}
{\mathbf u}(l)\propto\hat{\mathbf n} \exp(-\lambda t +i l q), \quad \hat{\mathbf n} \in \{\hat{\mathbf e}_1, \hat{\mathbf e}_2\} \,, \label{nmode-ansatz}
\end{align}
\noindent where $\lambda$ is the decay rate and $q \in [0,\pi]$ is the wave number. We note that $(E^{-m}-2+E^{m})u_{\mu}(l)=(e^{-imq}-2+e^{imq})u_{\mu}(l)$ and finally obtain the decay rates and effective spring constants $k^{||,\perp}$
\begin{align}
&\lambda^{||,\perp}=1+4 b^2 C^{||,\perp} \sum_{m=1}^{\infty} \frac{\sin^2(mq/2)}{m^5}\equiv k^{||,\perp}\,,
\label{keff-linear}
\end{align}
\noindent the result valid for subcritical fields, $b \le b_c$. The nearest neighbor (NN) approximation, where only one term with $m=1$ in sum \reff{keff-linear} is retained, works well and makes the result transparent. Relation \reff{keff-linear} is reduced to
\begin{align}
& \lambda_{NN}^{||}=1+4\left(\frac{b}{b_c^{NN}}\right)^2\sin^2 \left(\frac{q}{2}\right)\equiv k_{NN}^{||} \,,  \label{keff-long-NN}\\
& \lambda_{NN}^{\perp}=1-\left(\frac{b}{b_c^{NN}}\right)^2\sin^2\left(\frac{q}{2}\right) \equiv k_{NN}^{\perp}\,, \label{keff-perp-NN}
\end{align}
\noindent which allows us to draw a number of conclusions. First, this result shows hardening and softening of the spring constants with the field in the longitudinal and transverse directions, respectively. Second, it shows that as $b$ approaches its critical value $b_c$, the mode that first becomes unstable is the mode with the zigzag symmetry, $q=\pi$. Third, in a real system with a finite number of particles, the spectrum of decay rates is discrete. Its analysis helps in clarifying the impact of the chain length on the explosion patterns, as discussed in Sec.~\ref{ssec:expl-defects}. In Appendix~\ref{app:disprel-HI}, we also show how the dispersion relation \reff{keff-linear} can be generalized for the case of hydrodynamic interactions, when all particles are globally coupled through the solvent.

\subsection{Chain of finite length at finite temperature} \label{ssec:norm-modes-fin-kT}

Although highly enlightening, the analysis performed in Sec.~\ref{ssec:norm-modes-analytics} is based on small perturbations and corresponds to the harmonic approximation ($\alpha =0$) for trapping potential \reff{gauss-trap-pot}. To analyze softer, anharmonic potentials ($\alpha>0$) and achieve quantitative agreement between the experiment and theory, we apply a complementary approach which is well suited from both experimental and numerical perspectives.

Experimentally, the phonon-dispersion relations were determined from the single-particle trajectories obtained by video-microscopy \cite{Keim-etal-04}. We measured particle displacements ${\mathbf u}(l)$ from their equilibrium positions ${\mathbf R}(l)$. The Fourier transforms of the displacement vectors $u_{\mu}(q)=N^{-1/2}\sum_n u_\mu(n)\exp(-iqn)$ are directly related to harmonic potential energy \reff{U-harm-gen} as
\begin{equation}
U^{harm}=\frac{1}{2}\sum_{q}\sum_{\mu, \nu} u_\mu^*(q) D_{\mu \nu}(q) u_{\nu}(q)
\label{U-harm-fourier} \nonumber
\end{equation}
with $D_{\mu\nu}(q)$ the dynamical matrix \cite{Ashcroft-Mermin}. Applying equipartition, which implies that every mode has an energy of $k_BT/2$, leads to a relation \cite{Keim-etal-04}
\begin{equation}
\left < u_\mu^*(q) u_{\nu}(q) \right > = k_BT\, D_{\mu \nu}^{-1}(q),
\label{dynmatr-rel}
\end{equation}
\noindent where the average is over all independent configurations. The left-hand side of Eq.~\reff{dynmatr-rel} is accessible in both the experiment and numerical simulations, whereas the eigenvalues of $D_{\mu\nu}(q)$ yield the normal mode spring constants.

In a similar manner, the spring constants can be extracted from Brownian dynamics (BD) simulations. The dynamics of each particle is determined by the force ${\mathbf f}$ comprising deterministic, ${\mathbf f}_d$, and stochastic, ${\mathbf f}_s$, contributions. The deterministic counterpart ${\mathbf f}_d$ can be obtained from the energy of the colloidal crystal, Eq.~\reff{tot-energy}.  The impact of thermal fluctuations is modeled by Gaussian white noise. As a result, the dimensionless Langevin equation governing the dynamics of particle $l$ can be written as
%
\begin{align}
& \dot {\mathbf r}(l)={\mathbf f}(l), \quad {\mathbf f}(l)={\mathbf f}_d(l)+ \mathbf{f}_s(l) \, \label{langevin-eq} \\
& {\mathbf f}_d(l)= -\frac{\partial U}{\partial {\mathbf r}(l)} = 3b^2\sum_{l'\ne l}\frac{\mathbf{r}_{ll'}}{r^5_{ll'}}
-\delta\mathbf{r}_l\exp\left[-\frac{\alpha}{2}\delta\mathbf{r}^2_l\right]\,, \label{f-d} \\
& \left<{\mathbf f}_s(l,t)\right>=0, \quad \left<{\mathbf f}_s(l,t) {\mathbf f}_s(l',t')\right>=2\sigma \,{\mathbf I}\,\delta_{ll'}\delta(t-t')\,, \label{f-s}
\end{align}
%
\noindent where ${\mathbf r}_{ll'}={\mathbf r}(l)-{\mathbf r}(l')$, $r_{ll'}=|{\mathbf r}_{ll'}|$, ${\mathbf I}$ is the identity tensor of the second order, and $\delta_{ll'}$ and $\delta(t-t')$ being, respectively, Kronecker's and Dirac's delta functions. The parameters $\alpha$ and $b$ are defined by Eqs.~\reff{alpha-def} and \reff{b2-def}, respectively. The intensity of thermal fluctuations is determined by the dimensionless parameter
\begin{align}
\sigma = \frac{k_B T}{k_0d^2}\,, \label{epsilon-def}
\end{align}
\noindent which represents the thermal energy relative to the chosen energy scale. Note that the fact that we work at a given temperature $T$, and at certain values of $d$ and $k_0$, fixes the value of $\sigma$.

\begin{figure}[!h]
\includegraphics[width=0.42\textwidth]{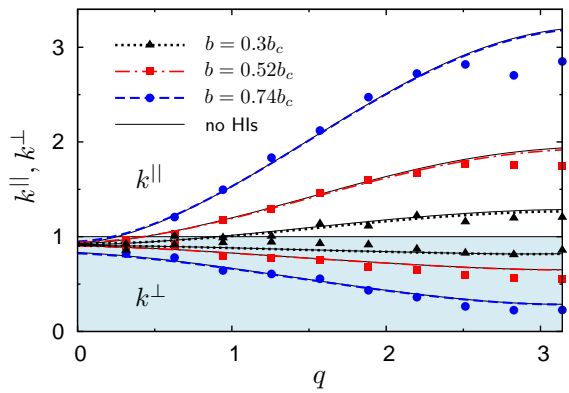}
\caption{(Color online) Comparison of the longitudinal $k^{||}$ and transverse $k^{\perp}$ normal mode spring constants obtained from the experiment (markers) and the BD simulations (lines) for different magnetic fields as a function of the wave number $q$. The bold dotted, dot-dashed, and dashed lines are from BD simulations with the HI for $\alpha=30$. The solid lines show the corresponding curves from the BD simulation without HI.} \label{fig:phonon-disp-rel}
\end{figure}

Figure \ref{fig:phonon-disp-rel} shows the phonon-dispersion relations as a function of $q$ for different values of the magnetic field $b<b_c$, which manifests good {\em quantitative} agreement between the experiment and BD simulations. To fit these data, we tuned $b$ and $\alpha$ such that a pair of $k^{||}(q)$ and $k^{\perp}(q)$ measured for the same magnetic field are in correspondence simultaneously, which eventually leads to correspondence for all pairs of $k^{||}(q)$ and $k^{\perp}(q)$ at $\alpha \simeq 30 \pm 5$. We note that the variation of parameter $\alpha$ affects the transverse and longitudinal modes differently, that is why quantitative agreement is not achievable via result \reff{keff-linear}, which was derived for strictly harmonic trapping, $\alpha=0$.

Note that in contrast to formula \reff{keff-linear}, which predicts $k^{||,\perp}(q=0)=1$, for $\alpha \ne 0$ we obtain at $q=0$: $k^{||}\ne k^{\perp}$, and both $k^{||,\perp} <1$. The fact that $k^{||,\perp} <1$ follows directly from the anharmonic nature of the trap, as shown explicitly in Appendix~\ref{app:nonharm-pot} for the vanishing magnetic field, $b=0$. However, the point that $k^{||}\ne k^{\perp}$ is further based on the fact that we work at the magnetic field $b >0$, which has different impacts on the longitudinal and transverse modes.

Finally, we note that the Langevin equations can be generalized to estimate the influence of hydrodynamic interactions (HI). In this case, instead of Eq.~\reff{langevin-eq} one simulates Eq.~\reff{mobility-rel}, in which $\mathbf{f}(l')=\mathbf{f}_d(l')+\mathbf{f}_s(l')$ is the total force acting on particle $l'$ with the deterministic force given by Eq.~\reff{f-d} and the generalized relations $\left<{\mathbf f}_s(l,t)\right>=0$, $\left<{\mathbf f}_s(l,t) {\mathbf f}_{s}(l',t')\right>=2\sigma \,{\mathbf H}^{-1}_{ll'}\delta(t-t')$ for the stochastic force. Here, ${\mathbf H}_{ll'}$ is the Oseen tensor introduced in Appendix~\ref{app:disprel-HI}, see expression \reff{H-tensor}. To numerically integrate the Langevin equations, both with and without hydrodynamic interactions, we applied a standard algorithm \cite{Ermak-McCammon-78}. To ensure that the colloids do not overlap in simulations, we have additionally included in the Langevin equations steric repulsive interactions of the Weeks-Chandler-Andersen form \cite{Weeks-Chandler-Andersen-71}. For spring constants, we have found very similar results obtained via BD simulations with and without HI, see Fig.~\ref{fig:phonon-disp-rel}. Note that in the simulations the HI were taken into account in the simplest form that neglects the existence of the boundary. The real HI will be modulated by the surface, see Appendix \ref{app:disprel-HI}, but given the small overall effect of HI, we can argue that explicitly including surface effects is not important for determining the spring constants.

\section{Nonequilibrium pattern formation. Dynamically expanding zigzag pattern}\label{sec:dynamic-zigzag}

\subsection{Infinite chain, no thermal noise} \label{ssec:expl-ideal}

To gain insight into the nonequilibrium process triggered by switching off the optical traps, we first consider an infinite chain of colloids in the limit of vanishing thermal noise. The positions of beads corresponding to the zigzag configuration are described by the vector ${\mathbf r}(l)={\mathbf h}(l)$ as given by Eq.~\reff{zigstate-h} with the transversal displacement now being a function of time, $h=h(t)$. The motion of particle $l$ satisfies the equation ${\mathbf v}(l)=\dot{\mathbf h}(l)={\mathbf f}_d(l)$. The force ${\mathbf f}_d(l)$ is given by expression \reff{f-d} evaluated at positions ${\mathbf h}(l)$, in which we retain the term $\propto b^2$ made by repulsive interactions but skip the contribution caused by the optical trapping. Because of symmetry, the longitudinal components of the velocity and force on all particles are vanishing and for the transversal ones we obtain
\begin{align}
& v^{\perp}(l) = f_d^{\perp}(l)=(-1)^l F^{\perp}(h)\,, \label{v-perp(l)} \\
& F^{\perp}(h)=6 b^2 \sum_{m=1}^{\infty}p_m h \,f_m^5(h)\,, \label{F-perp}
\end{align}
\noindent where $p_m$ and $f_m(h)$ in Eq.~\reff{F-perp} are given by Eq.~\reff{fm-pm-def}. By evaluating $v^{\perp}(l)=\hat{\mathbf e}_2\cdot\dot{\mathbf h}(l)=(-1)^l (\dot h/2)$ from Eq.~\reff{zigstate-h} and comparing the result with Eq.~\reff{v-perp(l)}, we arrive at the differential equation for $h=h(t)$
\begin{align}
& \frac{dh}{dt}=2 F^{\perp}(h)\,, \label{ode-h(t)}
\end{align}
\noindent which we supplement by the initial condition $h(t=0)=h_0>0$. Generally, Eq.~\reff{ode-h(t)} admits no analytical solution.

An explicit solution to Eq.~\reff{ode-h(t)}, however, can be obtained in the nearest-neighbor (NN) approximation, when only the term with $m=1$ is retained in the sum. By rescaling the time $\tau=12b^2 t$ and proceeding to a new variable $z=\sqrt{1+h^2}$, we arrive at the differential equation $z^6\dot z=z^2-1$, which admits an analytic solution
\begin{align}
&\frac{1}{5}z^5+\frac{1}{3}z^3+z-{\rm atanh}\left(\frac{1}{z}\right)=\tau+c\,, \label{sol-z(t)}
\end{align}
\noindent where $c$ is the integration constant determined by the initial condition at $t=0$: $z=z_0=\sqrt{1+h^2_0}$. The asymptotic solutions at small and large values of $h$ are given by expressions
\begin{align}
\ln\left(\frac{h}{h_0}\right)+\frac{5}{4}(h^2-h_0^2)=\tau \quad &(h_0 \le h \ll 1)\,, \label{asympt-small-h}\\
\frac{1}{5}h^5+\frac{5}{6}h^3=\tau \quad &(h_0 \ll h, \;h \gg 1)\,. \label{asympt-large-h}
\end{align}
\noindent Thus, at small times the displacement grows exponentially, $h(\tau)\approx h_0\exp(\tau)$, while at larger times the growth significantly slows down and then behaves according to a power law $h(\tau)\sim \tau^{1/5}$, independently of the initial conditions.

\begin{figure}[!h]
\includegraphics[width=0.45\textwidth]{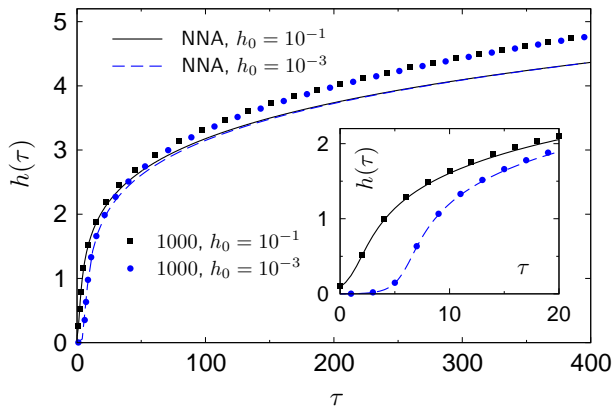}
\caption{(Color online) Transverse distance $h(\tau)$ for the infinite zigzag configuration as a function of the rescaled time $\tau$. The lines represent the solution in the NN-approximation (NNA), Eq.~\reff{sol-z(t)}, plotted for different initial conditions, $h_0=h(t=0)=0.1$ (solid line) and $h_0=0.001$ (dashed line) in units of lattice spacing $d$. The markers show the results of numerical integration of Eq.~\reff{ode-h(t)} for $1000$ neighboring particles the same initial conditions as for the approximate solution. The inset shows the solutions at the initial stage of evolution.} \label{fig:h-expl}
\end{figure}

Figure \ref{fig:h-expl} shows a comparison of the analytical solution \reff{sol-z(t)} valid in the NN-approximation with the numerical solution of Eq.~\reff{ode-h(t)} that accounts for interactions with $1000$ neighboring particles. If we double the number of neighboring particles, it does not change the results. The NN-approximation works very well, showing only a slight deviation from the numerical solution at large times $\tau$. At small distances $h$ (small time $\tau$), the contribution made by the interaction with the nearest neighbors is dominant. At large distances (large time $\tau$), when $h$ becomes larger than the period of the array, the interactions with more neighboring particles makes some small additional contribution. The reason is that at large $h$, the distances to the nearest and the next nearest neighbors are no longer drastically different.

\subsection{Impact of thermal noise}\label{ssec:expl-th-noise}

The nonequilibrium pattern formation at the vanishing temperature considered in Sec.~\ref{ssec:expl-ideal} is an idealization that has to be modified in the presence of thermal fluctuations. Since the repulsive interactions are long ranged, in an unbounded domain without thermal fluctuations this nonequilibrium process never stops and the particles move out to infinity. In reality, at some point thermal fluctuation that lead to diffusive behavior will start to dominate. This means that the nonequilibrium pattern can be characterized by a maximal spatial extension that is achieved within a certain time. To estimate these characteristics, we keep focussing on the infinite chain of particles and work in the NN-approximation.

A natural dimensionless measure that characterizes the relative strength of deterministic and diffusive motion of a particle is the P{\'e}clet number, ${\rm Pe}=D/(a v)$, where $D=k_BT/\xi$ is the coefficient of diffusion, $v$ is the characteristic deterministic velocity of the particle, and $a$ is the particle radius. For our dynamic configuration in the form of nonequilibrium zigzag, the absolute value of this velocity follows from Eqs.~\reff{v-perp(l)} and \reff{F-perp}. In the NN-approximation, we have
\begin{align}
& v (h) = 6b^2 h (1+h^2)^{-5/2}\,, \label{char-detvel}
\end{align}
\noindent which is strongly dependent on the separation distance $h$. At small $h$, the velocity grows linearly with $h$, $v \simeq 6b^2 h$, then the growth stops and at $h=1/2$ the velocity reaches the maximum, $v_{m}=96 \sqrt{5}b^2/125$. Afterwards, $v(h)$ starts to decay with $h$ and at large values of $h$ drops as $v \simeq 6b^2 h^{-4}$.

Thus, in the beginning of the nonequilibrium process, in the experiments the deterministic motion dominates, with the maximum value ${\rm Pe} = a v_m/D >1$, while at large times, when the motion becomes practically purely diffusive, ${\rm Pe} <1$. Therefore, the characteristic time can be defined as the time to reach a regime of motion with a small enough velocity such that ${\rm Pe}\approx 1$. In dimensional units this corresponds to the condition $v \approx k_B T/(a\xi)$. Recalling that the velocity is measured in the scales of $k_0 d/\xi$ and accounting for Eq.~\reff{char-detvel}, this condition yields the equation for the maximal distance $h_m$
\begin{align}
& h_m (1+h_m^2)^{-5/2} = \kappa, \quad \kappa = \frac{\sigma}{6b^2}\frac{d}{a} \,, \label{expl-hm}
\end{align}
\noindent where the parameters $b$ and $\sigma$ are given by Eqs.~\reff{b2-def} and \reff{epsilon-def}, respectively. A reasonable approximation for $h_m$ follows from Eq.~\reff{expl-hm} considered at large $h_m$, which provides the typical transversal displacement
\begin{align}
& h_m = \kappa^{-1/4} \,. \label{expl-hm-appr}
\end{align}

As follows from Eq.~\reff{expl-hm-appr}, for a given lattice period $d$, the distance $h_m$ grows with the increase in the strength of repulsive interactions and with the decrease in temperature. In the limit of no thermal noise, $k_B T \to 0$, we obtain $h_m \to \infty$, which is consistent with the findings of Sec.~\ref{ssec:expl-ideal}. For our experimental system ($B_0=1.9\,{\rm mT}$, $T=20^{\,\circ}{\rm C}$), we have $\kappa \approx 0.02$, which leads to $h_m \approx 2.6$ in lattice units, according to the approximate formula \reff{expl-hm-appr} or to the more exact estimate $h_m \approx 2.4$, as prescribed by Eq.~\reff{expl-hm}. Note that these estimates are in agreement with the trajectories obtained from the experiment and BD simulations, see Figs.~\ref{fig:explosion} and \ref{fig:chain-diff-N}.

As the motion of particles remains predominantly deterministic until $h$ reaches the value $h_m$, the time $\tau$ of the nonequilibrium expansion of the zigzag pattern can be estimated from Eq.~\reff{sol-z(t)} taken at $h=h_m$ or be alternatively read off directly from Fig.~\ref{fig:h-expl}. Note that as can be seen from Fig.~\ref{fig:h-expl}, the asymptotic solution for large $h$, see Eq.~\reff{asympt-large-h}, starts to work well already for $h \gtrsim 2.5$. Therefore, for the timescale of interest we obtain
\begin{align}
& \tau \approx \frac{1}{5} h_m^5 = \frac{1}{5}\kappa^{-5/4} \,. \label{expl-tau-appr}
\end{align}
%

\subsection{Impact of hydrodynamic interactions}\label{ssec:expl-HI}

\begin{figure}[!h]
\includegraphics[width=0.42\textwidth]{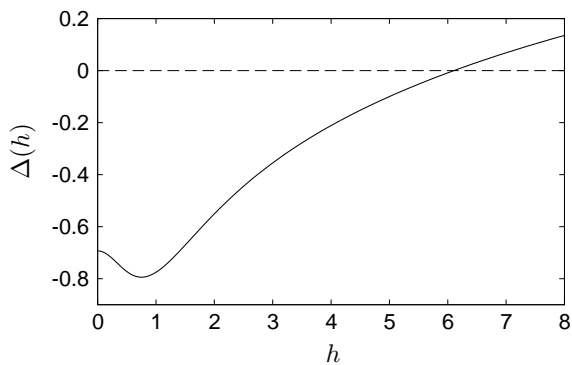}
\caption{The dependence $\Delta(h)$, as given by Eq.~\reff{Delta(h)} evaluated with $10000$ terms in the sum.} \label{fig:delta-h_hi}
\end{figure}

To understand the role of hydrodynamic interactions (HI), which as explained in Appendix~\ref{app:disprel-HI} are taken into account in the simplest form, we neglect thermal fluctuations and first focus on the infinite chain of particles, as in Sec.~\ref{ssec:expl-ideal}. Because of symmetry, the HI can change only the transversal component of velocity. In this case, according to representation \reff{mobility-rel}, Eq.~\reff{v-perp(l)} has to be replaced by $v_{\rm HI}^{\perp}(l)=\sum_{l'} H^{\perp}_{ll'}f^{\perp}_{d}(l')$. The sum over $l'$ is split into the term with $l'=l$ and the subsums with $l'=l \pm n$, $n=1, 2, \dots\,$. Taking into account that $f_d^{\perp}(l\pm n)=(-1)^n v^{\perp}(l)$ and evaluating the necessary components of the Oseen tensor
\begin{align}
& H^{\perp}_{ll} =1, \quad H^{\perp}_{ll \pm n}=\frac{3}{4}\frac{a}{d}\Delta_n(h)\,, \nonumber \\
& \Delta_n(h)=f_n(h)\left[ 1+ p_n h^2 f_n^2(h)\right]\,  \nonumber 
\end{align}
\noindent with $p_n$ and $f_n(h)$ defined by Eq.~\reff{fm-pm-def}, we obtain the velocity of beads modified by hydrodynamic interactions
\begin{align}
& v_{\rm HI}^{\perp}(l) = v^{\perp}(l)\left[ 1 + \frac{3}{2} \frac{a}{d} \Delta(h) \right] \,, \label{v-perp-hi} \\
& \Delta(h) = \sum_{n=1}^{\infty}(-1)^n \Delta_n(h) \label{Delta(h)} \,.
\end{align}
\noindent Here, $v^{\perp}(l)$ is the velocity of particle $l$ in the absence of HI and the hydrodynamic correction to velocity is multiplicative and the effect is governed by the small parameter $a/d$. For small $h$ we can evaluate expression \reff{Delta(h)}, which yields
\begin{align}
& h \ll 1: \quad \Delta(h) = - \ln 2 -\frac{7}{16}\zeta(3) h^2 +O(h^4) \,.
\end{align}
\noindent Note that in the limit of vanishing $h$ we have $\Delta(0) = - \ln 2$, Eq.~\reff{v-perp-hi} is reduced to
\begin{align}
& h\to 0: \quad v_{\rm HI}^{\perp}(l) \to v^{\perp}(l)\left( 1 - \frac{3}{2} \frac{a}{d} \ln 2 \right)\,,  \label{v-perp-hi-h=0}
\end{align}
\noindent and the overall hydrodynamic correction coincides with that obtained for the zigzag mode for $\lambda^{\perp}$, cf. Eq.~\reff{keff-HI-zmode}. The fact that the function $\Delta(h)$ is negative is a signature that HI effectively slow down the nonequilibrium process. As follows from Fig.~\ref{fig:delta-h_hi}, the function $\Delta(h)$ remains negative at distances $h \le 6$, which covers practically the whole range of distances not masked by thermal fluctuations. As argued in Sec.~\ref{ssec:expl-th-noise}, this range corresponds to the distances $h \lesssim 2.5$. This means that the interpretation about HI as a factor equivalent to the effective slowdown of velocities remains valid for all distances of interest.

\begin{figure}[!h]
\includegraphics[width=0.46\textwidth]{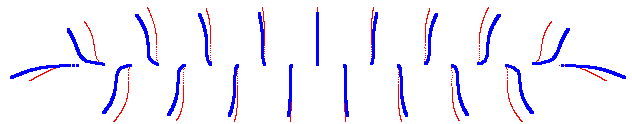}
\caption{(Color online) Comparison of trajectories for a chain of $N=19$ particles with (bold blue lines) and without (thinner red lines) hydrodynamic interactions, obtained by BD simulations with no thermal noise, $\sigma=0$, and $b=0.8 b_c$ for the same time interval of $\Delta\tau=125$. The initial state corresponds to the zigzag configuration with a small transversal displacement $h_0=0.1$. An effective  slowdown of the expansion speed becomes evident from the trajectory of the central particle, which has traveled a shorter distance due to HI.} \label{fig:expl-hi}
\end{figure}

Up to now we have considered the infinite configuration of particles. Now it is easy to predict what happens in a chain with a finite number of particles. First, consider the case of no HI. Because the chain is no more infinite, the longitudinal component of repulsive forces is compensated only for particles in the middle of the chain. For particles which have significantly different numbers of neighbors on the left and on the right, this force is nonvanishing. As a result, the particles acquire the longitudinal component of velocity and their trajectories bend from those in the case of infinite chain with the ideally transverse motion. This effect of deflection is maximal for the particles at the ends of the chain, while the transversal component of motion is relatively weak, and the particles move away from the chain, see Fig.~\ref{fig:expl-hi}.

Consider now the impact of HI on this nonequilibrium process. On the one hand, as follows from the consideration of the infinite configuration without HI, the transverse velocity component of particles is maximal at distances $h=1/2$ and then decays with $h$ as $h^{-4}$. On the other hand, the hydrodynamic correction to the velocity is generally small and lowers the instantaneous velocity. Taking into account these two facts, we can conclude that HI have maximal effect in the beginning of expansion process. The slowdown of velocities is, to some extent, equivalent to effectively stronger repulsive forces between the particles. As a result, the particles trajectories resemble the effect of a slightly stronger repulsive interactions in the beginning of the nonequilibrium process, which leads to slightly more pronounced bending of trajectories compared to the case of no HI, see Fig.~\ref{fig:expl-hi}.

\subsection{General case. Defects in nonequilibrium patterns} \label{ssec:expl-defects}

The normal mode analysis performed in Sec.~\ref{ssec:norm-modes-analytics} for an infinite chain of particles shows that the mode that first becomes unstable is the zigzag mode, the fact that underlies the physics of nonequilibrium patterns. In contrast to this analysis, in a real system with a finite number of particles, the spectrum is discrete. The observed nonequilibrium patterns do not always have the perfect zigzag symmetry and can involve defects. Before proceeding to BD simulations demonstrating these features, we consider a simplified model that reveals the origin of the defects.

Consider a chain of finite number $N$ of particles, which can be odd, $N=2M+1$, or even, $N=2M$. For convenience, we choose the origin of the coordinate system in the middle of the pattern such that ideal zigzag patterns would be symmetric for odd $N$ and antisymmetric for even $N$. In other words, we assume that the dimensionless coordinates of laser traps are given by ${\mathbf R}(l)=(l,0)$, $l=0, \pm 1, \pm 2, \dots, \pm M$ for an odd $N$ and by ${\mathbf R}(l)=(l \mp 1/2,0)$, $l=\pm 1, \pm 2, \dots, \pm M$ for an even $N$ and estimate the normal modes from Eq.~\reff{nmode-ansatz}. As a result, we obtain
\begin{align}
& u_{\mu}^{(k)}(l) \propto e^{-\lambda_k t}\cos q_k l \;\; (l=0, \pm 1, \dots, \pm M)\,, \nonumber \\
& q_k=\frac{2\pi k}{N-1}, \;\; k=0,1,\dots,M \;\;\; (N=2M+1) \label{nmode-Nodd}
\end{align}
\noindent for an odd number $N$ of colloids and
\begin{align}
& u_{\mu}^{(k)}(l) \propto e^{-\lambda_k t}\sin q_k (l \mp 1/2), \;\; (l=\pm 1, \dots, \pm M)\,, \nonumber\\
& q_k=\frac{(2k-1)\pi}{N-1}, \;\; k=1,2,\dots,M \;\;\; (N=2M) \label{nmode-Neven}
\end{align}
\noindent for an even number of colloids. The spectrum of decay rates can be obtained from the NN-approximation, following Eqs.~\reff{keff-long-NN} and \reff{keff-perp-NN}
\begin{align}
& \lambda^{||}_k=1+\frac{4 b^2}{b_c^2}\sin^2 \left(\frac{q_k}{2}\right),  \;\;
\lambda^{\perp}_k=1-\frac{b^2}{b_c^2}\sin^2\left(\frac{q_k}{2}\right)\,.  \label{decrate-discr}
\end{align}

As earlier, the zigzag mode has the shortest spatial period or, equivalently, the largest wave number, $k=M$, $q_M=\pi$, see Eqs.~\reff{nmode-Nodd} and \reff{nmode-Neven}, giving $u_{\mu}^{(M)}(l) \propto \cos \pi l$ and $u_{\mu}^{(M)}(l) \propto \sin[ \pi (l \mp 1/2)]$. As follows from relation \reff{decrate-discr}, the zigzag mode has the slowest transverse decay rate, see Eq.~\reff{trans-decrat} for $\lambda_M^{\perp}$, and first becomes unstable as we approach the critical field $b=b_c$. Note that other modes have close decay rates and may lead to defects, if the zigzag mode is not well separated from the others.

It is instructive to analyze what happens as the length of the chain is changed. If for the zigzag mode, $q_M=\pi$, the wave number for other modes $k=M-m$, $m=1, 2, \dots$ close to the zigzag mode is, as follows from Eqs.~\reff{nmode-Nodd} and \reff{nmode-Neven}, $q_{M-m}=q_M-\beta_m \pi$. Here, $\beta_m=m/M$ for the case of odd number of colloids, $N=2M+1$, and $\beta_m=(2m+1)/(2M+1)$ for the case of even number of colloids, $N=2M$. The transversal decay rates can be represented as
\begin{align}
& \lambda^{\perp}_{M} = 1- \frac{b^2}{b_c^2}, \;\; \lambda^{\perp}_{M-m}=\lambda^{\perp}_{M}+\frac{b^2}{b_c^2}\sin^2\left(\frac{\beta_m\pi}{2}\right)\,.  \label{trans-decrat}
\end{align}

By inspecting relation \reff{trans-decrat}, we can draw an important conclusion. In the case of small fields, significantly below the threshold, $b \ll b_c$, all the modes have the transverse decay rates $\lambda^{\perp}_{M} \approx \lambda^{\perp}_{M-m}\approx 1$. As the repulsive interactions are relatively weak and the modes are damped, we do not observe well pronounced zigzag symmetry in the nonequilibrium patterns, which is additionally masked by thermal fluctuations. The situation becomes qualitatively different as we approach the threshold.  For the fields close to but remaining below the threshold, $b \lesssim b_c$, we obtain from Eq.~\reff{trans-decrat}
\begin{align}
& \lambda^{\perp}_{M} \approx 0, \quad \lambda^{\perp}_{M-m} \approx \sin^2\left(\frac{\beta_m\pi}{2}\right)\,.  \label{trans-def}
\end{align}
\noindent Relation \reff{trans-def} implies that the modes are best separated provided $\lambda^{\perp}_{M-n}$ has a maximal value, which is achieved at $\beta_m=1$. By definition, $\beta_m \le 1$, and the modes are better separated for larger values of $\beta_m$. If we consider the mode closest to the zigzag ($m=1$), then we see that $\beta_m$ becomes larger for smaller $M$ and smaller for larger $M$. In other words, the neighboring modes close to the zigzag mode are better separated from each other for shorter chains ($N$ small) and worse separated for longer chains ($N$ large).

\begin{figure}[!h]
\includegraphics[width=0.42\textwidth]{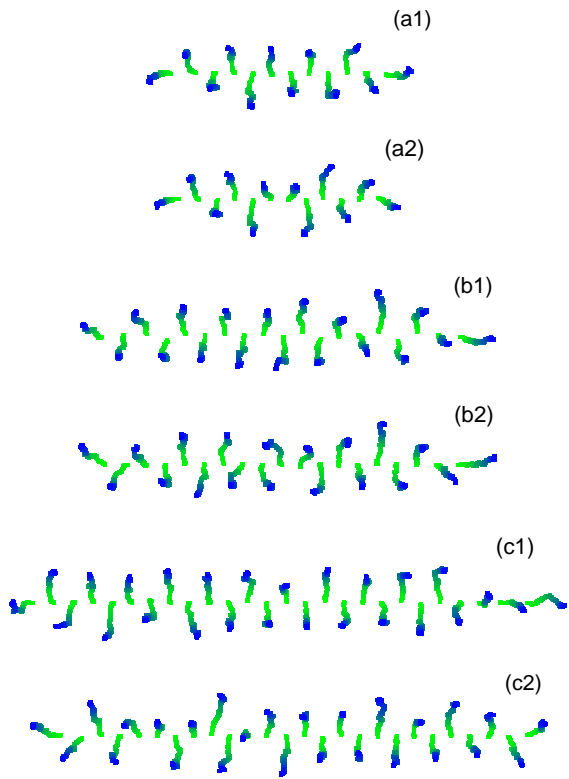}
\caption{(Color online) Particle trajectories showing nonequilibrium expansion patterns for chains with different number of colloids: $N=12$ (a), $N=19$ (b), and $N=26$ (c) obtained for $b=0.8 b_c$ and $\alpha=30$. Panels (a1), (b1), and (c1) demonstrate zigzag patterns without defects, while panels (a2), (b2), and (c2) show patterns with simple imperfections. } \label{fig:chain-diff-N}
\end{figure}

For a system at a finite temperature, the probability to detect a mode different from the zigzag mode grows with the chain length, which results in defects, as confirmed by our BD simulations, Fig.~\ref{fig:chain-diff-N}. For instance, the mode $k=M-1$ closest to the zigzag has a slightly larger period than the zigzag mode. Being considered within the same length, this implies the appearance of defects. The simplest defect corresponds to the situation when a pair of neighboring particles ``shoots out'' in the same direction, breaking the perfect zigzag symmetry, see panels (a2), (b2), and (c2) of Fig.~\ref{fig:chain-diff-N}. Another defect is composed of three neighboring particles, with the outer particles moving in the opposite directions and nearly resting particle between them, as in panel (c2) of Fig.~\ref{fig:chain-diff-N}, cf. Fig.~1(b) of Ref.~\cite{Pyka-etal-natcomm-13}.

\section{Conclusions}\label{sec:conclusions}

We study the behavior of repulsively interacting paramagnetic colloidal particles which are initially optically trapped in a linear array and form a nonequilibrium expanding pattern when the traps are abruptly turned off. This dynamical pattern exhibits a zigzag symmetry even when the strength of the magnetic interactions, characterized by the dimensionless parameter $b$, is weaker than the critical value $b=b_c$  required to break the linear symmetry of the equilibrium state imposed by the optical traps. Theory and computer simulations quantitatively replicate these phenomena both in and out of equilibrium.

An analysis of equilibrium states shows that the line state is always stable for $b < b_c$. For harmonic traps, the transition to the zigzag state occurs at $b=b_c$ and this equilibrium state is stable for $b>b_c$. For anharmonic traps, specified by a dimensionless softness parameter $\alpha$, there exists a critical value $\alpha_c$ that separates two qualitatively different scenarios. For relatively soft traps, $\alpha < \alpha_c$, the transition to the zigzag state is still possible. However, in contrast to the case of harmonic traps ($\alpha=0$), the zigzag state remains stable only in a finite range of fields, $b_c < b < b_{**}$. For $b > b_{**}$ the zigzag configuration becomes unstable and the zigzag pattern starts to dynamically expand. For stiffer traps, $\alpha > \alpha_c$, the transition to the equilibrium zigzag state is impossible. The line state loses its stability and exhibits an expanding pattern upon reaching the value $b=b_c$. The thermal fluctuations are shown to effectively decrease the critical values $b_c$ (and $b_{**})$, at which the line (and zigzag) state becomes unstable. The corresponding corrections are additive and $\propto k_BT/V_0$, the small ratio of thermal energy ($k_BT$) to the depth ($V_0$) of the trap potential.

The normal mode analysis of the line configuration performed for both zero and nonzero temperatures evidences hardening and softening of the spring constants with the field in the longitudinal and transverse directions, respectively. This analysis further reveals that as $b$ approaches its critical value $b_c$, the mode that first becomes unstable is the mode with the zigzag symmetry, which explains the symmetry of nonequilibrium patterns. We demonstrated that at zero temperature, the particle is unable to explore the anharmonic nature of the trap potential and the spring constants correspond to the purely harmonic trap. The anharmonic corrections, again governed by the small parameter $k_BT/V_0$, are negative, leading to effective spring constants smaller than those for the harmonic trap and meaning that the anharmonic trap potential is softer than its purely harmonic counterpart. The theoretically predicted phonon dispersion relations are in good quantitative agreement with the experimental data. From this comparison we could also draw an estimate for the softness parameter specifying our system, $\alpha > \alpha_c$. In agreement with the experiment, this value means that the equilibrium zigzag state is impossible for our system. We have also found that hydrodynamic interactions have only a small effect on the results of the normal modes analysis.

We have developed a description explaining the formation and evolution of nonequilibrium zigzag patterns. The basic physics is captured by a simple model for an infinite chain of particles in the absence of thermal fluctuations. In the nearest-neighbor approximation, we obtain an analytic solution for the trajectories of particles in the chain. We show that accounting for the magnetic interactions with further neighbors does not significantly change the results. Further we demonstrate that in contrast to the idealized case of no temperature, where the process of expansion formally never stops, we obtain an estimate for a characteristic distance at which thermal fluctuations start to dominate and lead to diffusive behavior. This distance is found to be a few lattice periods, in agreement with the experiment. The analysis of chains of finite length shows that the trajectories of particles at the ends of the chain start to bend. In the beginning of expansion, the hydrodynamic interactions effectively are shown to effectively slow down the velocity of expansion, which eventually results in a more pronounced bending of the trajectories. Finally, considering chains of finite length in the presence of thermal fluctuations and taking into account the analytical results of the normal mode analysis we explain the existence of defects in zigzag patterns and illustrate why the defects become more probable in longer chains.

\acknowledgements
AVS was supported by HPC-EUROPA2 (Project No. 228398) and Volkswagen Foundation (project I/83903), which are gratefully acknowledged.

\appendix

\section{Correction to dispersion relation caused by hydrodynamic interactions}\label{app:disprel-HI}

Generally, the motion of driven particles is coupled through the solvent, which can be taken into account within the concept of hydrodynamic interactions (HI). In the presently considered case of overdamped motion, the dynamics of a given particle, say, particle $l$, depends on the motion of other particles and satisfies the equation
\begin{align}
& {\mathbf v}(l)=\dot {\mathbf r}(l)=\sum_{l'}{\mathbf H}_{ll'}\cdot{\mathbf f}(l')\,, \label{mobility-rel}
\end{align}
\noindent where ${\mathbf H}_{ll'}$ is the mobility tensor and ${\mathbf f}(l')$ is the force exerted on particle $l'$. In the Oseen approximation, valid at interparticle distances large compared to the particle size and implying no confinement, the dimensionless mobility tensor can be written as
\begin{align}
& {\mathbf H}_{ll'}=\delta_{ll'}\, {\mathbf I} + (1-\delta_{ll'}) \frac{3}{4}\frac{a}{d}\frac{1}{r_{ll'}}\left( {\mathbf I}
+ \hat{\mathbf r}_{ll'} \otimes \hat{\mathbf r}_{ll'} \right)\, \label{H-tensor}
\end{align}
\noindent with $\hat{\mathbf r}_{ll'}={\mathbf r}_{ll'}/r_{ll'}$.

Note that in the presence of a confining surface, as in our experiment, the mobility tensor given by Eq.~\reff{H-tensor} has to be replaced by the Blake tensor \cite{Blake-71}, which accounts for the confinement. Such an approach can be applied at the level of both the Oseen \cite{Wollin-Stark-11} or a more accurate Rotne-Prager approximation \cite{Netz-etal-06}. For the discussion of further improvements such as, e.g., many-body and lubrication effects, see Refs.~\cite{Jones-Kutteh-99, Swan-Brady-07, Swan-thesis-10}. However, in the BD simulations for spring constants, where the mobility tensor was taken into account in the simplest form that neglects the existence of the boundary, as in Eq.~\reff{H-tensor}, we have found very similar results with and without HI, see Fig.~\ref{fig:phonon-disp-rel}. Given the small overall effect of HI, we can argue that explicitly including surface effects is not important for determining the spring constants and therefore restrict ourselves to the Oseen approximation.

Now, to obtain the correction to the dispersion relation due to HI, Eq.~\reff{NM-eq-umu} for particle displacements from the laser traps, which eventually determines spring constants, has to be modified according to representation \reff{mobility-rel}. For $l=l'$ the mobility tensor \reff{H-tensor} has the simplest structure, ${\mathbf H}_{ll'}={\mathbf I}$, as in the absence of HI. For $l \ne l'$, we use small particle displacements $u_{\mu}(l)$ in comparison with the period of array $d$. As a result, we obtain approximate expressions
\begin{align}
& {\mathbf H}_{ll'}\approx\left(
\begin{array}{cc}
{H}_{ll'}^{||} & 0 \\
0 & {H}_{ll'}^{\perp}
\end{array}
\right)\,,  \\
& {H}_{ll'}^{||} = 2 {H}_{ll'}^{\perp} \approx \frac{3}{2}\frac{a}{d}\frac{1}{|l-l'|} \quad (l \ne l') \,, \label{}
\end{align}
\noindent within the same accuracy as the Oseen approximation.

According to representation \reff{mobility-rel}, Eq.~\reff{NM-eq-umu} has to be replaced by $\mathbf{\dot{u}}(l) = \sum_{l'} \mathbf{H}_{ll'} \mathcal L \mathbf{u}(l)$ with the operator $\mathcal{L}$ given by Eq.~\reff{LinOper-L}. Splitting the sum over $l'$ into the term with $l'=l$ and two subsums with $l'=l \pm n$ ($n=1, 2, \dots $), we arrive at the generalized equation for perturbations that accounts for HI
\begin{align}
&\dot{u}_{\mu}(l) =  \mathcal L_{\mu}^{\rm HI} \mathcal L_{\mu} u_{\mu}(l) \,, \label{NM-eq-umu-HI} \\
& \mathcal L_{\mu}^{\rm HI}= 1 + \frac{3}{4} \frac{a}{d} c_{\mu} \sum_{n=1}^{\infty} \frac{E^{-n}+E^{n}}{n}\,, \label{LinOper-LHI}
\end{align}
\noindent where the coefficients $c^{||} \equiv c_1 =2$ and $c^{\perp} \equiv c_2 =1$. Applying the same ansatz for $u_{\mu}$ as earlier, see Eq.~\reff{nmode-ansatz}, and taking into account that $\sum_{n=1}^\infty n^{-1}\exp(\pm i n q) =-\ln[1-\exp(\pm iq)]$, we arrive at the generalized expressions for the spring constants
\begin{align}
&\lambda^{||}_{\rm HI}=\lambda^{||}\left[1-3 \frac{a}{d} \ln \left( 2\sin \frac{q}{2}\right)\right]\equiv k^{||}_{\rm HI}\,, \label{keff-long-HI} \\
&\lambda^{\perp}_{\rm HI}=\lambda^{\perp}\left[1-\frac{3}{2} \frac{a}{d} \ln \left( 2\sin \frac{q}{2}\right)\right]\equiv k^{\perp}_{\rm HI}\,,
\label{keff-perp-HI}
\end{align}
\noindent with $\lambda^{||}$ and $\lambda^{\perp}$ defined by general relation \reff{keff-linear} or its NN-approximation, see Eqs.~\reff{keff-long-NN} and \reff{keff-perp-NN}. In the case of no HI, when the ratio $a/d$ is formally set to zero, the generalized expressions reduce to those obtained earlier, $\lambda^{||,\perp}_{\rm HI}=\lambda^{||,\perp}$. In the partial case of no repulsive interactions, when $b=0$ and $\lambda^{||}=\lambda^{\perp}=1$, expressions \reff{keff-long-HI} and \reff{keff-perp-HI} are in agreement with those obtained in by Polin {\it et al.} \cite{Polin-etal-06}.

Based on the structure of the generalized spring constants we can draw the following conclusions: i) The hydrodynamic correction is small, being of order $O(a/d)$, which reflects the accuracy the Oseen approximation; ii) For the zigzag mode, $q=\pi$, we find $\lambda^{||}_{\rm HI}=\lambda^{||}[1-(3a/d)\ln 2]$ and
\begin{align}
& \lambda^{\perp}_{\rm HI}=\lambda^{\perp}\left(1-\frac{3}{2} \frac{a}{d} \ln 2\right)\,, \label{keff-HI-zmode}
\end{align}
\noindent
which show destabilizing role of hydrodynamic interactions close to the critical point that determines the transition to the equilibrium zigzag state; iii) The hydrodynamic correction is multiplicative and hence the critical point itself is not affected by HI. The latter point is not unexpected because HI refer to dynamics, while the transition to the equilibrium zigzag state is a purely equilibrium feature.

\section{Particle in a weakly anharmonic potential}\label{app:nonharm-pot}

To qualitatively demonstrate the role of the anharmonic nature of the trapping potential we address a simplified problem. Consider the one-dimensional motion of a single particle subject to thermal fluctuations and trapped by a potential of form \reff{gauss-trap-pot} in the absence of magnetic field ($b=0$), which presents a partial case of the problem described by Eqs.~\reff{langevin-eq}-\reff{f-s}. Denoting by $x$ the displacement from the laser trap, the Langevin equation is reduced to
\begin{align}
& \dot x = f_d + f_s, \quad f_d = -\frac{\partial V_T}{\partial x}\,, \label{lang-1d-eq} \\
& \left<f_s(t)\right>=0, \quad \left<f_s(t),f_s(t')\right>=2\sigma \,\delta(t-t')\, \label{lang-1d-sf}
\end{align}
\noindent with the trap potential $V_T=\alpha^{-1} \left[1-\exp\left(-\alpha x^2/2\right)\right]$, in which the parameter $\alpha$ is a measure of anharmonicity of the trapping potential, see Eq.~\reff{alpha-def}. Further we consider the case of the weakly anharmonic trap, $\alpha \ll 1$. Using the smallness of $\alpha$, we expand to find
\begin{align}
& V_T(x) = \frac{1}{2}x^2 + \frac{1}{8}\alpha x^4 + \mathcal{O}(\alpha^2)\,, \label{lang-1d-VT}
\end{align}
\noindent where the first term corresponds to the limit of a purely harmonic trap, $\alpha=0$, and the second one accounts for the anharmonicity. This representation can be interpreted as a nonlinear restoring force $f_d(x)$ with a coordinate-dependent stiffness $K(x)$
\begin{align}
& f_d = -K(x) x, \quad K(x)=1-\frac{1}{2} \alpha x^2 \,. \label{lang-1d-K(x)}
\end{align}

To obtain the effective spring constant, the local stiffness $K(x)$ has to be weighted with the probability of finding the particle at the coordinate $x$ and integrated all over the domain, $-\infty < x < \infty$. The general stationary solution for the probability density function $P_0(x)$ for problem \reff{lang-1d-eq}, \reff{lang-1d-sf} is known to be \cite{Risken-book-96}
\begin{align}
& P_0(x) = \mathcal{C}^{-1} \exp\left(-\frac{V_T}{\sigma}\right), \quad \mathcal{C}=\int_{-\infty}^{\infty} \exp\left(-\frac{V_T}{\sigma}\right) dx \,. \nonumber
\end{align}
\noindent For a weakly anharmonic potential \reff{lang-1d-VT}, we obtain a consistent approximation
\begin{align}
& P_0(x) \approx \mathcal{C}^{-1} \exp\left(-\frac{x^2}{2\sigma}\right)\left(1+\frac{1}{8}\frac{\alpha}{\sigma}x^2\right)\,, \nonumber \\
& \mathcal{C} \approx \sqrt{2\pi\sigma}\left(1+\frac{3}{8}\alpha \sigma\right) \,. \nonumber
\end{align}
Using this result, we evaluate the effective spring constant $k = \left<K(x)\right> = 1 - \alpha \sigma/2 +\mathcal{O}(\alpha^2\sigma^2)$, where $\left<\dots\right>=\int_{-\infty}^{\infty}\dots P_0(x)dx$. Interestingly, this rigorous result coincides with the expression following from averaging Eq.~\reff{lang-1d-K(x)} for $K(x)$, if we heuristically put $\left<x^2\right>=\sigma$ based on the purely Gaussian probability density function, which is our $P_0(x)$ taken for the harmonic potential, $\alpha=0$.

Finally, by using the definitions of $\alpha$ and $\sigma$, we formulate the result for the effective spring constant in the dimensional form to yield
\begin{align}
& k = k_0\left( 1 - \frac{1}{2}\frac{k_B T}{V_0}\right) +\mathcal{O}\left(\frac{k_B^2T^2}{V_0^2}\right)\,. \label{lang-1d-keff}
\end{align}

We see that in the case of vanishingly small thermal fluctuations, $k_BT=0$, the particle is unable to explore the anharmonic nature of the trap potential and the spring constant corresponds to the purely harmonic trap, $k=k_0$. The anharmonic correction is governed by the small parameter $\alpha\sigma = k_BT/V_0$,
whose structure implies that the deviation from $k_0$ becomes visible in the case of thermal noise provided the trap potential has a finite depth. The sign of the correction tells us that the effective spring constant is smaller than $k_0$, which simply reflects the fact that the trap potential is softer than its purely harmonic counterpart.


\end{document}